\documentclass[sigconf]{acmart}

\usepackage{graphicx}
\usepackage{xcolor}
\usepackage{multirow}
\usepackage{hyperref}
\usepackage{enumitem}

\usepackage{makecell}
\usepackage{framed}

\definecolor{shadecolor}{gray}{0.95}

\copyrightyear{2026}
\acmYear{2026}
\setcopyright{cc}
\setcctype{by}
\acmConference[ICSE '26]{2026 IEEE/ACM 48th International Conference on Software Engineering}{April 12--18, 2026}{Rio de Janeiro, Brazil}
\acmBooktitle{2026 IEEE/ACM 48th International Conference on Software Engineering (ICSE '26), April 12--18, 2026, Rio de Janeiro, Brazil}
\acmDOI{10.1145/3744916.3787813}
\acmISBN{979-8-4007-2025-3/2026/04}

\newboolean{showcomments}
\setboolean{showcomments}{true} % Change to false to hide comments
\ifthenelse{\boolean{showcomments}}
{% If showcomments is true
\newcommand{\nbc}[3]{
\colorbox{#3}{\bfseries\sffamily\scriptsize\textcolor{white}{#1}}
{\textcolor{#3}{\sf\small$\blacktriangleright$\textit{#2}$\blacktriangleleft$}}
}
}
{% If showcomments is false
\newcommand{\nbc}[3]{}
}

\begin{document}

\title{The State of Open Science in Software Engineering Research: A Case Study of ICSE Artifacts}

\author{Al Muttakin}
\affiliation{%
  \institution{University of Saskatchewan, Canada}
  \city{}
  \state{}
  \country{}
}
\email{al.muttakin@usask.ca}

\author{Saikat Mondal}
\affiliation{%
  \institution{University of Saskatchewan, Canada}
  \city{}
  \state{}
  \country{}
}
\email{saikat.mondal@usask.ca}

\author{Chanchal K. Roy}
\affiliation{%
  \institution{University of Saskatchewan, Canada}
  \city{}
  \state{}
  \country{}
}
\email{chanchal.roy@usask.ca}

\renewcommand{\shortauthors}{Muttakin, et al.}

\begin{abstract}
Replication packages are crucial for enabling transparency, validation, and reuse in software engineering (SE) research. While artifact sharing is now a standard practice and even expected at premier SE venues such as ICSE, the practical usability of these replication packages remain underexplored.
In particular, there is a marked lack of studies that comprehensively examine the executability and reproducibility of replication packages in SE research.
In this paper, we aim to fill this gap by evaluating 100 replication packages published in ICSE proceedings over the past decade (2015--2024).
We assess the (1) executability of the replication packages, (2) efforts and modifications required to execute them, (3) challenges that prevent executability, and (4) reproducibility of the original findings for those that are executable.
We spent approximately 650 person-hours in total to execute the artifacts and reproduce the study findings.
Our analysis shows that only 40 of the 100 evaluated artifacts were fully executable. Among these, 32.5\% ran without any modification. However, even executable artifacts required varying levels of effort: 17.5\% required low effort, while 82.5\% required moderate to high effort to execute successfully.
We identified five common types of modifications and 13 challenges that lead to execution failure, encompassing environmental, documentation, and structural issues. Among the executable artifacts, only 35\% (14 out of 40) reproduced the original results.
These findings highlight a notable gap between artifact availability, executability, and reproducibility. 
Our study proposes three actionable guidelines to improve the preparation, documentation, and review of research artifacts, thereby strengthening the rigor and sustainability of open science practices in SE research.
\end{abstract}




\maketitle

\section{Introduction}
\label{sec:introduction}

Open science promotes transparency, accessibility, and reusability by encouraging researchers to share publications, data, code, and methods with the community \cite{Mendez2020OpenScience}. Backed by mandates from governments, funding organizations, and publishers, open science
has gained strong policy and institutional support and is widely promoted for enabling reproducibility, transparency, and more efficient research practices \cite{OECD2015, RoyalSociety2012, Mirowski2018, Pontika2015}.
Software engineering (SE) research is highly empirical and tool-centric, making it naturally well-suited for open science practices. However, the adoption of open science in SE has remained slow and inconsistent 
\cite{Mendez2020OpenScience}. In recent years, leading SE conferences and journals have taken steps to promote open science by introducing policies and artifact evaluation tracks, offering dedicated reviews and badges to incentivize the sharing of data, tools, and code \cite{timperley2021}. A recent study shows encouraging trends: 67.7\% of SE papers from top-tier conferences (e.g., ICSE, FSE, ASE) between 2017 and 2022 included artifacts \cite{Liu2024}. However, their actual usability in terms of executability and reproducibility, which are
core aspects of open science, remains uncertain. When curated effectively, these artifacts can enhance trust, enable reuse, and support rigorous validation and comparison across studies. Thus, a systematic investigation of the state of open science in SE is warranted to assess current practices, identify challenges to executability and reproducibility, and guide future improvements.

Several studies have explored open science in SE by examining artifact metadata, conducting surveys, reviewing the literature, and performing methodological analyses \cite{Solari2018LabPackages, Cordeiro2025a, Mahmood2018, Gundersen2018}. For example, 
\citet{Solari2018LabPackages} analyzed seven lab packages and proposed a reference model for artifact structure, highlighting common 
structural patterns. 
\citet{Cordeiro2025a} surveyed SE researchers to understand reproducibility practices in controlled experiments. 
\citet{RodriguezPerez2018} systematically reviewed 187 SZZ-related studies, which assessed reproducibility and reported their limitations. In a broader context, 
\citet{Gundersen2018} analyzed 400 AI papers to evaluate the availability of source code and the quality of documentation required for reproducibility. However, they did not attempt to execute or reproduce the studies. Furthermore, 
\citet{Nong2023} evaluated 11 tools across 55 papers, but focused narrowly on deep learning-based vulnerability detection in C/C++.
While these efforts offer valuable insights, they primarily rely on secondary sources, with limited direct engagement in the execution and evaluation of artifacts. 
Therefore, a comprehensive and hands-on assessment of open science practices across the broader SE research landscape remains an open challenge.

In this study, we present a comprehensive empirical assessment of open science practices in SE by analyzing the \textit{executability} and \textit{reproducibility} of 100 replication packages from International Conference on Software Engineering (ICSE) papers published over the past decade. First, we collected all ICSE research-track papers published from 2015 to 2024 and systematically inspected each paper to determine whether it included a link to a replication package. For those who included a link, we visited it to see if it is still alive and accessible. We then reviewed the contents of each accessible package to determine whether it included executable components. Second, from the set of packages with executable components, we selected 100 replication packages using stratified random sampling \cite{koyuncu2009ratio} and executed them in an isolated environment to systematically assess the degree of success and the effort required to achieve it.
Third, for successfully executed packages, we compared the obtained results with those reported in the original studies to assess reproducibility. We spent approximately 650 person-hours executing the replication packages and analyzing their reproducibility. In particular, we made four major contributions by answering the following four research questions.

$\bullet$ \textbf{RQ\textsubscript{1} (\textit{Executability, Effort \& Modification}): 
To what extent are replication packages executable, and what level of effort and modifications are required to execute them?} 
Executability is essential for validating, reusing, and extending research findings. Non-functional artifacts offer limited value and undermine scientific progress.
We executed 100 ICSE replication packages in a controlled environment, strictly following the original authors' documentation.
Execution outcomes were labeled as executable, partially executable, or not executable. For each artifact, we recorded the required effort, categorized as low, moderate, or high, and grouped the necessary modifications into five dimensions (e.g., environment setup, code updates).

$\bullet$ \textbf{RQ\textsubscript{2} (\textit{Types of Modifications}): What types of modifications are needed to execute replication packages?}
Identifying the nature and frequency of required modifications reveals recurring quality issues in artifact packaging and helps guide best practices. We thus attempt to execute replication packages manually and document all required modifications. These were organized into a hierarchical taxonomy comprising five broad categories and 14 specific types.

$\bullet$ \textbf{RQ\textsubscript{3} (\textit{Execution Barriers}): What challenges prevent the successful execution of replication packages?}
Identifying common failure points helps improve artifact design and review standards. For artifacts that failed or were partially executable, we documented the challenges that prevented execution
and categorized them into three broader themes (e.g., documentation gaps) and 13 specific types, along with their frequencies.

$\bullet$ \textbf{RQ\textsubscript{4} (\textit{Reproducibility}): Do executable replication packages reproduce the original results, and what factors contribute to reproducibility failure?}
Reproducibility is critical for verifying empirical results and ensuring the credibility of research claims. We evaluated 64 artifacts that were either fully or partially executable and categorized reproduction outcomes as fully reproducible, partially reproducible, not reproducible, unverifiable, and no output generated. We also examined factors that contribute to reproducibility failures.

\smallskip
\noindent Full \textbf{Replication Package} of our study can be found in our online appendix \cite{ReplicationPackage}.

\begin{figure}[!htb]
\centering
\includegraphics[width=3.4in]{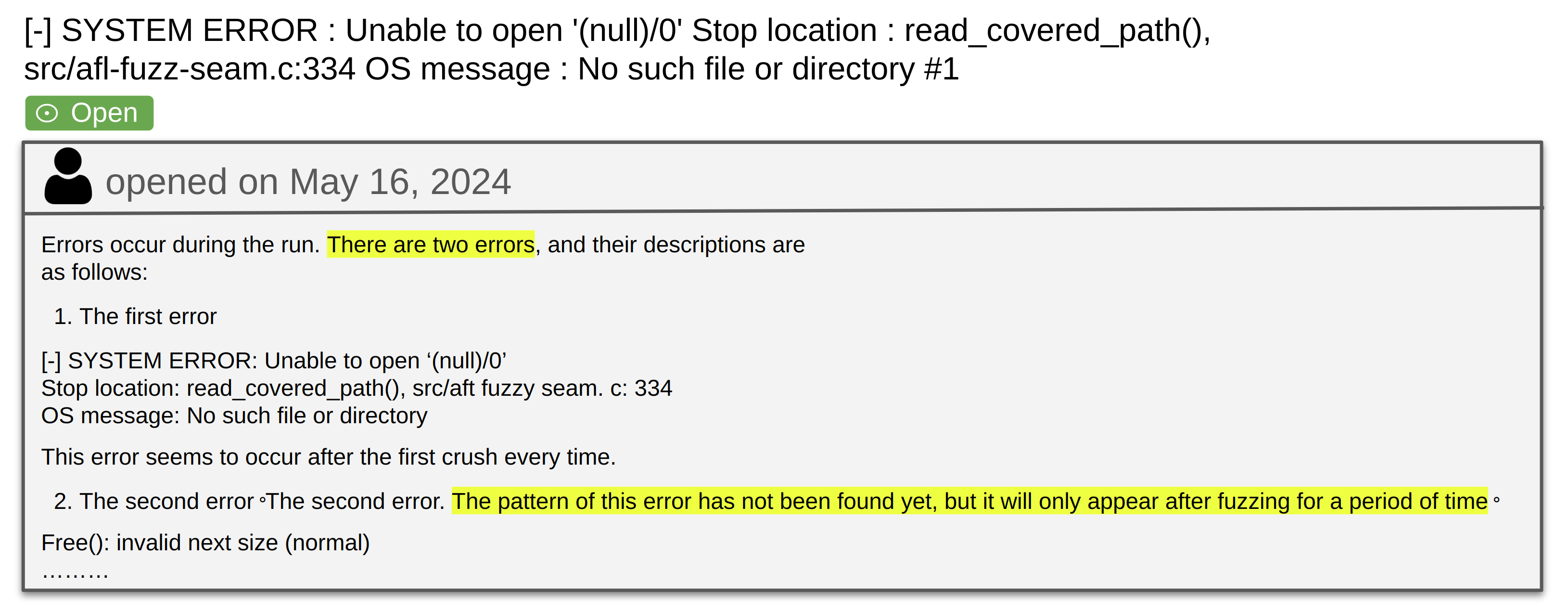}
\caption{Example of an issue \cite{seamfuzz_issue1} in a replication package, where users encountered critical execution errors.}
\Description{A screenshot of a GitHub issue showing error messages and stack traces encountered when attempting to execute a replication package.}
\label{fig:ME1}
\end{figure}

\section{Motivating Example}
\label{sec:motivating-examples}

\begin{figure}[htb]
\centering
\includegraphics[width=3.2in]{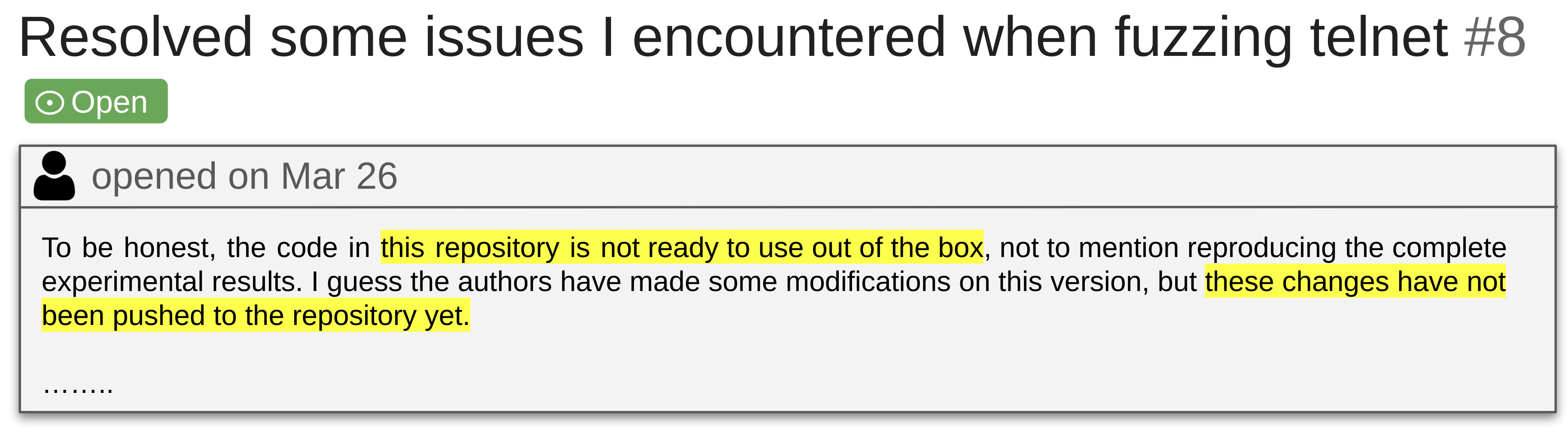}
\caption{Example of a user's concern about the quality of the artifact \cite{ltlfuzzer_issue8}.}
\Description{A screenshot showing user concerns about artifact quality, including comments about incomplete documentation, missing repository updates, and difficulty understanding how to execute the replication package.}
\label{fig:ME2}
\end{figure}

Despite growing efforts to promote open science, a notable gap exists between the availability of replication packages and their practical usability (e.g., executability and reproducibility). Consider two real-world examples that illustrate this gap. Figure~\ref{fig:ME1} illustrates an executability issue in a replication package. 
In particular, it highlights an issue reported in the replication package uploaded to GitHub for an ICSE 2023 paper \cite{Lee2023SEAMFUZZ}, where a user attempted to run the shared artifact but encountered critical errors, potentially due to missing files or undocumented environmental assumptions.
Notably, this artifact received both the \textit{available} and \textit{reusable} badges \cite{acm_artifact_review_2024}, indicating that existing artifact evaluation criteria may not always capture certain practical usability challenges.
Figure~\ref{fig:ME2} presents another case from the ICSE artifact~\cite{Meng2022LTTFuzz}, where a user 
reported that the code was not usable out-of-the-box due to missing required updates that had not been pushed to the public repository. As a result, the replication attempt was unsuccessful despite the artifact being publicly available and endorsed.

In both cases, the publicly shared artifacts were not executable due to undocumented issues, underscoring the gap between availability and practical reusability. These examples illustrate that while sharing an artifact might reflect an intent to support open science, incomplete or non-executable artifacts can limit their value, slow down SE research, and reduce the impact of the original work.

Motivated by these observations, we present a large-scale empirical assessment of 100 ICSE artifacts to examine their executability and reproducibility. Our findings uncover executability challenges, highlight best practices, and provide actionable recommendations to strengthen the state of open science in SE.

\begin{table*}[t]
\centering
\caption{Summary of ICSE Papers, Artifact Availability, and Executability (2015-2024)}
\label{tab:artifact-summary}
\resizebox{6.5in}{!}{%
\begin{tabular}{lccccccccccc}
\hline
\textbf{Metric} & \textbf{2015} & \textbf{2016} & \textbf{2017} & \textbf{2018} & \textbf{2019} & \textbf{2020} & \textbf{2021} & \textbf{2022} & \textbf{2023} & \textbf{2024} & \textbf{Total} \\
\hline
\hline

\# of Papers              & 84  & 101 & 68  & 105 & 109 & 129 & 138 & 197 & 207 & 234 & \textbf{1372} \\

\hline

Artifact Links & 47 (55.95\%) & 51 (50.50\%) & 39 (57.35\%) & 60 (57.14\%) & 75 (68.81\%) & 98 (75.97\%) & 121 (87.68\%) & 175 (88.83\%) & 197 (95.17\%) & 213 (91.03\%) & \textbf{1085 (79.08\%)} \\

\hline

Available Artifacts & 17 (36.17\%) & 19 (37.25\%) & 30 (76.92\%) & 49 (81.67\%) & 68 (90.67\%) & 90 (91.84\%) & 111 (91.74\%) & 170 (97.14\%) & 187 (94.92\%) & 200 (93.90\%) & \textbf{941 (86.73\%)} \\

\hline
Executable Component & 12 (70.59\%) & 12 (63.16\%) & 24 (80.00\%) & 40 (81.63\%) & 59 (86.76\%) & 70 (77.78\%) & 89 (80.18\%) & 150 (88.24\%) & 167 (89.30\%) & 173 (86.50\%) & \textbf{796 (84.59\%)} \\

\hline

Stratified Sample          & 2   & 2   & 3   & 5   & 7   & 9   & 11  & 19  & 20  & 22  & \textbf{100} \\
\hline
\end{tabular}
}
\end{table*}

\section{ Study Methodology}
\label{sec:study-methodology}

\subsection{Dataset Construction}
\label{subsec:dataset-construction}

To evaluate open science practices in SE, we constructed a curated dataset of ICSE research papers and their associated replication artifacts spanning the past ten years (2015--2024). The following sections outline our rationale, paper collection strategy, artifact identification process, and availability assessment.

\noindent\textbf{Venue Selection Rationale.} Software engineering research encompasses a diverse set of subdomains and publication venues. Conducting a comprehensive evaluation of open science practices across the entire SE landscape would be infeasible. To ensure both representativeness and focus, we selected ICSE as the target venue for our study. As a premier conference in the field, ICSE consistently publishes high-quality technical research across a diverse range of SE topics, making it an ideal venue for evaluating artifacts, their executability, and reproducibility at scale.

\noindent\textbf{Corpus Collection.}
We collected 1,372 ICSE research track papers published from 2015 to 2024 from publication databases (e.g., IEEE, ACM) using their Digital Object Identifiers.
Table~\ref{tab:artifact-summary} shows the yearly distribution of ICSE papers with replication packages, accessibility, and executability readiness.

\noindent\textbf{Artifact Identification Process.} To determine whether a paper shared replication package links, we employed the following three-pass search strategy. In \underline{Pass 1}, we searched the PDF for open-science-related keywords (e.g., replication, reproduce, artifact, package, dataset) to identify artifact links. To construct this keyword list, we manually examined an initial set of 20 papers, which resulted in a list of 25 terms (please refer to our replication package \cite{ReplicationPackage}) 
used consistently throughout the analysis.
If no link was identified via keyword search, in \underline{Pass 2},  we manually scanned the full text, focusing on the end of the introduction, dedicated sections (e.g., Data Availability), footnotes, and the reference list to locate any implicit mentions of artifacts or URLs. 
For papers that discussed replication packages but did not include explicit artifact links, in \underline{Pass 3} we conducted web searches using the paper title and author names to determine whether artifacts were available on platforms such as GitHub, Zenodo, or institutional websites.
Interestingly, we found one artifact link in this process.
Among 1,372 ICSE papers, 1,085 (79.08\%) included at least one artifact link. While some artifacts were hosted on multiple platforms (21 cases), we collected all provided links to assess their availability.

\noindent\textbf{Artifact Availability with Executable Components.}
For papers that included artifact links, we applied two filtering criteria:
(1) \underline{availability} -- the provided links were active, and the artifact contents could be accessed at the time of inspection, and (2) \underline{executability} -- the artifacts contained executable components (e.g., code, scripts, or tools) to initiate an execution attempt.
Based on these filtering criteria, we identified 941 papers with available artifact links and 796 papers whose artifacts were truly available and had executable components. Among 1372 papers, 287 did not provide any replication package link, and 144 provided links that were inaccessible.

\noindent\textbf{Sampling Strategy.}
To ensure temporal representativeness, we applied year-wise stratified random sampling \cite{koyuncu2009ratio} to select 100 artifacts for execution and reproducibility analysis, as summarized in Table~\ref{tab:artifact-summary}. The choice of 100 packages balanced breadth and feasibility, ensuring wide coverage of SE studies while keeping the manual execution and assessment workload practical yet rigorous.
We initially planned to sample approximately 10\% from each year (around 80 packages), but expanded this to 100 to enhance diversity and strengthen the validity of our findings.
Although not based on a formal power analysis, this sample size is consistent with standard practice in empirical SE research, where extensive manual evaluation typically limits the proportion of artifacts analyzed while still supporting representative and reliable conclusions.

\subsection{Assessing Executability, Effort, and Required Modifications} %(RQ1 \& RQ2)
\label{subsec:executability-methodology}

To evaluate the executability of replication artifacts, we conducted a structured execution experiment involving two highly experienced professionals. The primary execution was conducted by the first author, who has over 12 years of software development experience across diverse technology stacks (e.g., Java, Python, C/C++, Docker) and previously served as a Software Technical Lead at a reputable software firm. This author spent approximately 500 person-hours executing 100 replication packages.
To ensure fair evaluation and reduce individual bias, the second author, with over 15 years of development experience, independently re-attempted execution for any package where the first author was unsuccessful, contributing an additional 150 person-hours.
Of the 37 artifacts that the first author could not execute, the second author re-ran all of them and changed the status of only one by identifying a missing execution parameter that enabled partial execution.
When both evaluators failed, either partially or entirely, we classified the artifact as \textit{Partially Executable} or \textit{Not Executable}, respectively.

We followed a two-step procedure for executing each artifact. First, we reviewed the documentation provided with the replication package to understand its functionality and intended execution flow. Documentation formats varied widely and were commonly provided through \texttt{README.md} files, plain text files bundled in the archive, or inline explanations embedded within code files (e.g., Jupyter Notebooks). For artifacts lacking documentation, we attempted to infer the execution logic directly from the source code. We imposed a two-hour time limit to bound our effort in such cases.
Once an execution strategy was identified, we proceeded with artifact setup and execution in a controlled local environment. The following sections detail the steps performed during execution.

\subsubsection{Retrieving and Preparing Artifacts}
Replication artifacts were hosted on various platforms, including Git-based repositories (e.g., GitHub), compressed archives hosted on research data repositories (e.g., Zenodo, FigShare), and project or institutional websites. Based on the hosting type, we either cloned the repository or downloaded and extracted the archive into a dedicated directory to prepare it for execution. 

\subsubsection{Setting up Isolated Execution Environments}
To ensure consistent and replicable execution, we prepared isolated environments tailored to each artifact's requirements, as discussed below.

$\bullet$ \textbf{\textit{Containerization and Virtual Environments}.}
We used Docker containers to simulate specific operating system (OS) versions and complex dependency stacks, particularly for artifacts requiring older Linux distributions. Conda environments were used to manage Python-related dependencies in a clean, reproducible manner. While several artifacts used their isolated environments (e.g., Dockerfiles or environment files), others required us to provision custom Docker images according to the documented requirements. 
This approach allowed us to emulate legacy setups without altering the host system and ensured consistency across executions. For example, one artifact from 2015 required Ubuntu 12.04. Since obtaining a machine running Ubuntu 12.04 in 2025 was impossible, we successfully replicated the environment using Docker.

$\bullet$ \textbf{\textit{Handling Missing Specifications}.}
Our primary source for system specifications was the artifact's documentation. However, when artifacts did not specify system requirements, we used our M2 machine (Table~\ref{tab:machines}) as the default hardware configuration. To approximate the OS version, we used a two-year backward buffer from the artifact's publication year, accounting for the typical gap between the start of the experiment and the publication date. For example, for an artifact published in 2020, we used the latest long-term support OS version available in 2018.

\subsubsection{Installing Dependencies and External Tools}
After setting up the isolated environments, we installed the dependencies and external tools required for execution as follows.

$\bullet$ \textbf{\textit{Following Provided Instructions}.}
We primarily relied on the artifact documentation to obtain dependencies. When installation instructions were provided, we followed them directly. If only a dependency list were available without installation instructions, we used our development experience and external resources to complete the setup.

$\bullet$ \textbf{\textit{Inferring Dependencies Without Documentation}.}
In many cases, artifacts lacked both installation instructions and explicit dependency lists. In these cases, we searched the artifact for language-specific dependency files. For Python artifacts, we installed Python and its package manager, then searched for files such as \texttt{require-}\\\texttt{ments.txt} or \texttt{environment.yml}.
We installed OpenJDK and Maven for Java artifacts and searched for files such as \texttt{pom.xml}. If no such files were present, we manually inspected import statements and installed dependencies one by one, a time-consuming and error-prone process. Unresolved dependencies at this stage were deferred to the troubleshooting phase. When version information was missing, we applied the two-year backward buffer heuristic, similar to the previous step, to estimate suitable versions.

\subsubsection{Running the Experiment}
\label{subsec:execution} 
After installing all dependencies, we attempted to execute the artifact. We first consulted the documentation to identify any executable commands provided by the authors. The type and granularity of instructions varied across artifacts. For example, studies involving Large Language Models often included commands for pretraining, fine-tuning, and evaluation. Others provided shell scripts to trigger an end-to-end execution pipeline or step-by-step instructions to run individual components.

However, many artifacts lacked execution instructions entirely. In such cases, we inspected the codebase to infer a plausible entry point for execution. The success of this process depended heavily on the technology stack and project structure. For instance, executing a standalone Python script is typically straightforward, unless undocumented runtime parameters are required. In more complex cases involving multiple scripts or loosely organized codebases, identifying the correct execution sequence without guidance was challenging.
Some programming languages follow standard conventions for entry points (e.g., \texttt{main.java} in Java projects), whereas others rely on user-defined conventions, such as naming scripts \texttt{main.py}, \texttt{run.py}, or \texttt{run.sh}. These cues often helped us infer where to begin, though no consistent standard was observed. In such cases, we relied on intuition, programming experience, and a structural understanding of the codebase.

Once a command was executed, we monitored the outcome. If it completed successfully, we proceeded to the next step. If it failed with a runtime error, we deferred to the troubleshooting phase.

\begin{table*}[!htb]
\centering
\caption{Machine Details}
\label{tab:machines}
\resizebox{6.5in}{!}{%
\begin{tabular}{cllcccl}
\hline
\textbf{ID} & \textbf{Operating System} & \textbf{Processor} & \textbf{\# Cores} & \textbf{RAM} & \textbf{Hard Drive} & \textbf{GPU}\\
\hline \hline
M1 & Windows 11, 64 bit & Intel® Core™ i7-13700H @ 3.7 GHz & 14 & 32 GB & 1 TB & Intel® Iris® Xe\\
\hline
M2 & Ubuntu 24.04 LTS, 64 bit & Intel(R) Core(TM) i7-12700K @ 3.6 GHz & 12 & 128 GB & 2 TB & NVIDIA GeForce RTX 3080 12 GB\\
\hline
M3 & Ubuntu 22.04.5 LTS, 64 bit & Intel(R) Xeon(R) Platinum 8356H CPU @ 3.90 GHz & 32 & 3 TB & 24 TB & 3 * NVIDIA A100 80 GB PCIe\\
\hline
M4 & Mac OS Sonoma 15.4 & Apple M3 Pro & 11 & 18 GB & 512 GB & 3 * 14 - Core built in GPU\\
\hline
M5 & Windows 11, 64 bit & Intel® Core™ i7-11700U @ 2.6 GHz & 4 & 32 GB & 1 TB & Intel® Iris® Xe\\
\hline
M6 & Ubuntu 22.04.5 LTS, 64 bit & Intel® Core™ i7-11700U @ 2.6 GHz & 4 & 32 GB & 1 TB & Intel® Iris® Xe\\
\hline
M7 & Ubuntu 22.04.5 LTS, 64 bit & Intel® Core™ i7-6700 @ 3.40 GHz & 4 & 32 GB & 1 TB & Intel Corporation HD Graphics 530\\
\hline
\end{tabular}%
}
\end{table*}

\subsubsection{Troubleshooting and Fixing Runtime Errors}
\label{subsub:modification-types}
If execution failed, we attempted to identify the root cause of the runtime error. Troubleshooting was guided by the evaluators' technical expertise, standard debugging practices, supporting resources (e.g., Stack Overflow, ChatGPT), and other online documentation. Once the issue was diagnosed, we explored and applied suitable fixes.
Depending on the type of error, we implemented a range of modifications to enable execution. Below, we summarize five high-level categories of fixes applied across different replication packages:

\textbf{ (i) Change Environment Setup.}
This category includes modifications such as installing missing dependencies, replacing unavailable or local packages (e.g., removing a Python package built from source with an official package from \texttt{pip} registry), resolving version conflicts, installing external tools, and updating system environment variables (e.g., \texttt{PATH}).

\textbf{(ii) Modify Instructions.}
In many cases, we modified the provided instructions by adjusting relative paths, correcting flag values, or adding missing commands that were not included in the documentation. For example, we changed the value of \texttt{-nproc\_per\_node} flag to match the number of GPUs in our environment.

\textbf{(iii) Source Code Modification}.
To resolve runtime errors, sometimes we made modifications to the artifact’s source code. This was often challenging, as poor or missing code documentation made it difficult to infer the intended logic and behavior (e.g., changing the input directory path in the source code to access the data file).

\textbf{(iv) Change File Organization.}
In many cases, execution failed due to incorrect file structure, missing files, or directories. We addressed these issues by reorganizing folders, creating missing directories, removing unnecessary directories, or reconstructing essential files required for execution.

\textbf{ (v) Adjust Configuration Files.}
We updated configuration files to align with our environment, including fixing directory paths and modifying control variables. For example, we updated the dataset path in \texttt{config.json} to enable the successful execution of the artifact.

Once modifications were applied, we returned to the previous execution step (Section~\ref{subsec:execution}) and re-ran the experiment. If the issues were resolved, we continued with the remaining execution steps. Otherwise, we resumed troubleshooting.
Troubleshooting was often time-consuming, with the number and complexity of errors varying across artifacts. To manage effort effectively, we set a maximum effort threshold of \textit{four} hours. If an issue could not be resolved within that window and the artifact lacked sufficient documentation, we stopped further investigation and proceeded to the final execution decision.

\subsubsection{Labeling}
This is the final stage of the execution process, where each artifact is assigned one of the following labels based on the execution and troubleshooting outcomes.

\begin{itemize}
    \item \textbf{Executable} -- able to complete the execution with or without modifications.
    \item \textbf{Partially Executable} -- able to execute some component of the artifact with or without modification.
    \item \textbf{Not Executable} -- cannot execute any component of the artifact following the given instructions and troubleshooting.
\end{itemize}

\smallskip
\noindent We further classified each of the above levels by effort level and required modification. Effort level is defined based on the total time spent attempting to execute the artifact:
\textit{Low Effort} -- less than 2 hours,
\textit{Moderate Effort} -- between 2 and 8 hours, and
\textit{High Effort} -- more than 8 hours.
Modification level reflects the time invested in troubleshooting and fixing issues before determining the final execution status:
\textit{Minor Modification} -- less than 1 hour,
\textit{Major Modification} -- between 1 and 4 hours, and
\textit{Critical Modification} -- more than 4 hours.
The challenges we encountered and the time required to complete the execution process were documented for further qualitative analysis. 

\subsection{Analyzing Failure Causes and Executability Challenges}
\label{subsec:challenges-methodology}

We examined artifacts that were non-executable or only partially executable to identify the underlying challenges hindering successful execution. The identified issues were then manually coded and organized into a structured taxonomy comprising several specific subgroups, which were further grouped into broader thematic categories based on their nature and frequency.

\subsection{Examining Reproducibility of Executable Artifacts and the Factors Behind Failure} %(RQ4)
\label{subsec:reproducibility-methodology}

We assessed reproducibility of results through a three-step process for all artifacts labeled as \textit{Executable} or \textit{Partially Executable}.
First, we documented the outputs generated during execution, whether partial or complete. 
Second, we checked whether the artifact included explicit validation instructions to compare the generated outputs with those reported in the original study. These instructions typically included one or more of the following:  (i) reference outputs or logs,  (ii) scripts or commands to compute performance metrics, or (iii) visualizations or result plots for comparison. When such validation steps were available, we followed them to verify whether our outputs matched or closely resembled the original results.
Finally, without validation guidance, we manually compared our outputs with results reported in the paper, focusing on the metrics, tables, and figures described. However, when the execution produced raw outputs that were not directly reported in the paper, and no scripts were provided to process or interpret them, we could not assess reproducibility.
Based on the outcome of this evaluation, we assigned each artifact to one of five reproducibility categories:

\begin{itemize}

    \item \textbf{Fully Reproducible} -- the artifact executed successfully and the results matched those reported in the original paper using either provided validation steps or manual comparison.

    \item \textbf{Partially Reproducible} -- the artifact executed successfully or partially, but only a subset of the outputs matched those reported in the original study.

    \item \textbf{Not Reproducible} -- the artifact executed successfully or partially, but none of the outputs matched the original study results, even after manual comparison.

    \item \textbf{Unverifiable} -- the artifact executed successfully or partially, but verification was not feasible due to missing validation instructions, reference results, or guidance on how to interpret the results.

    \item \textbf{No Output Generated} -- the artifact executed successfully or partially but produced no output.
    
\end{itemize}

% \smallskip
\noindent This systematic evaluation provides a large-scale empirical assessments of whether ICSE artifacts can actually reproduce the findings reported in their corresponding papers.

\subsection{Experimental Setup}
\label{sec:experimental-setup}

To conduct the execution and reproducibility experiments, the first two authors analyzed artifacts using a total of \textit{seven} machines, selected to closely align with the requirements specified in the replication packages. 
Table~\ref{tab:machines} provides an overview of the hardware and operating system specifications used in this study.
Each evaluator had access to at least \textit{one Windows machine} to run artifacts that provided Windows-specific instructions or executables, and to \textit{two Ubuntu-based Linux machines}, at least one of which was equipped with a GPU to support experiments involving hardware acceleration or deep learning workloads.
We kept a \textit{Mac machine} as a backup, and it was needed only for one artifact that required Mac-specific data processing.

\section{Study Findings}
\label{sec:findings}

This section presents the empirical findings for our four research questions, supported by data and evidence from our analysis.

\subsection{Answering RQ1: Executability Status, Level of Effort and Modifications to Execute Replication Package}
\label{subsec:findings-rq1}

\begin{figure}[htb]
\centering
\includegraphics[width=0.9\columnwidth]{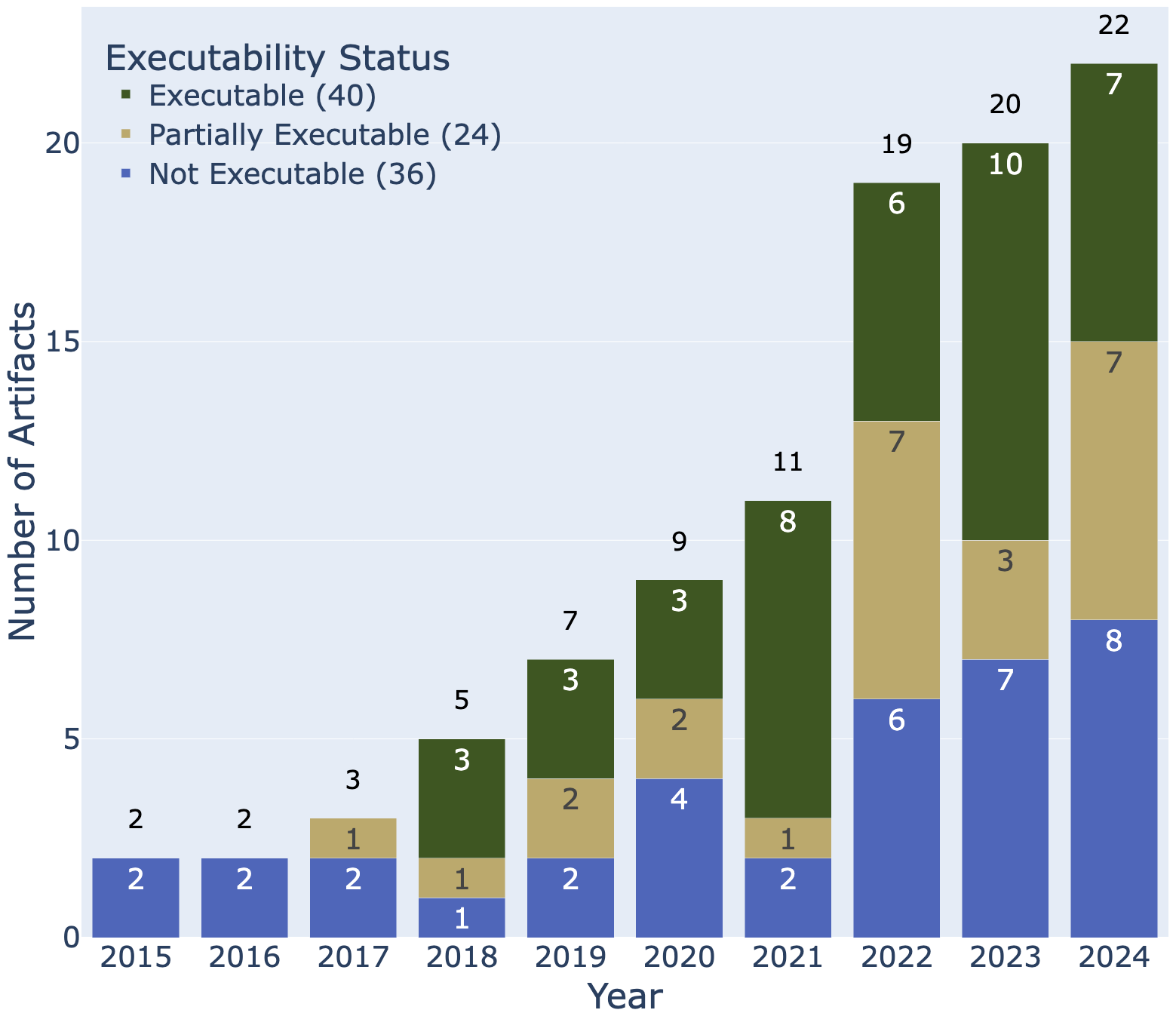}
\caption{Executability status of artifacts over the years.}
\Description{A stacked bar chart showing executability status from 2015 to 2024. Each bar represents one year and is divided into three segments: executable (green), partially executable (yellow), and not executable (red).}
\label{fig:RQ1_executability_by_year}
\end{figure}

\noindent\textbf{Executability of Replication Package.}
Figure~\ref{fig:RQ1_executability_by_year} shows the executability status of artifacts over the years. 
In 2024, only 7 of 22 artifacts (31.82\%) were fully executable, 8 (36.36\%) were not executable, and 7 (31.82\%) were partially executable. 
In contrast, 2023 shows a higher rate of full executability, with 10 of 20 artifacts (50\%) executing successfully, while 6 (30\%) failed completely and 4 (20\%) achieved only partial execution.
These figures suggest that recent artifacts are not consistently more executable despite increased availability.
The most promising year was 2021, with 8 out of 11 artifacts (72.73\%) fully executable, indicating a potential high point in artifact quality. However, this trend was not sustained in later years.
In earlier years (2015-2018), there were very few artifacts overall, which limited inference. Notably, all artifacts in 2015 and 2016 failed to execute.

\begin{figure}[t]
\centering
\includegraphics[width=3.4in]{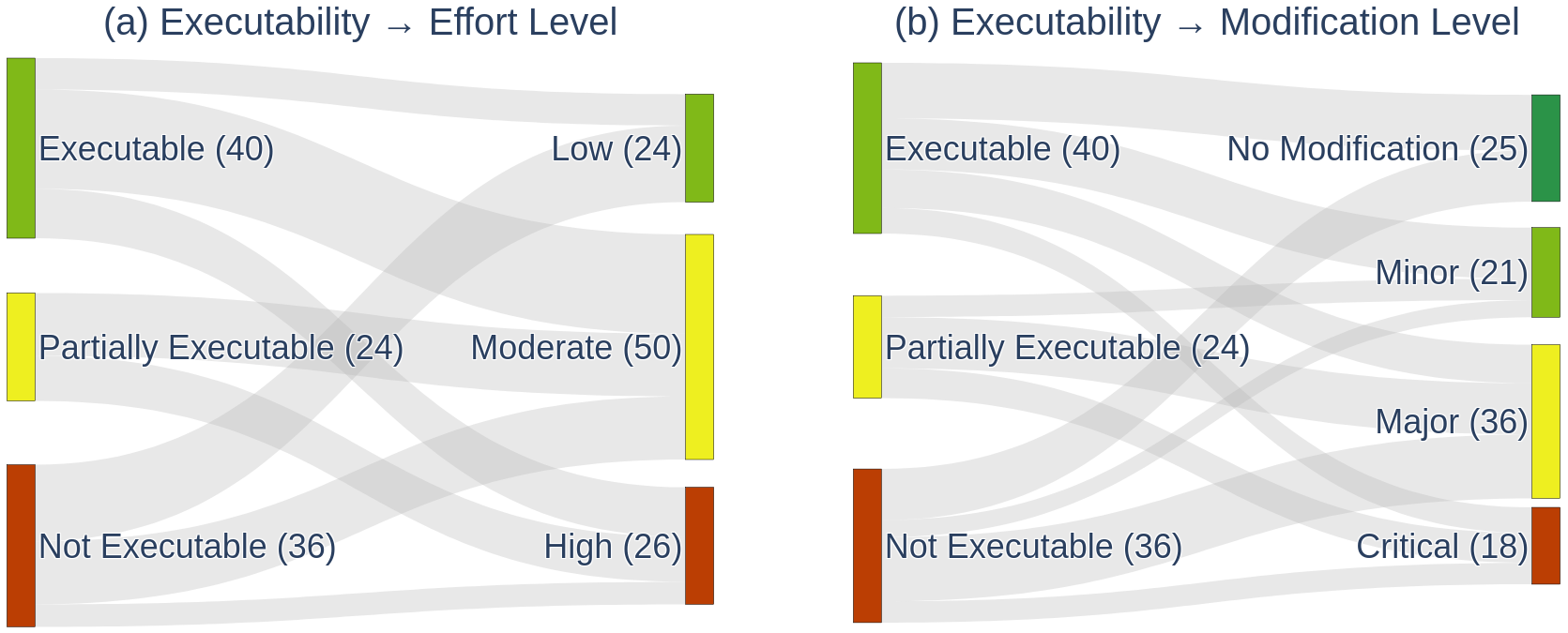}
\caption{Effort \& Modification Vs. Executability}
\Description{Two side-by-side stacked bar charts. The left chart shows effort levels (High, Medium, Low) required for execution. The right chart shows modification levels (Significant, Moderate, Minor, None) needed to make artifacts executable.}
\label{fig:RQ1_effort_mod}
\vspace{-3mm}
\end{figure}

In total, across all years, 40 out of 100 artifacts (40\%) were executable, 24 (24\%) partially executable, and 36 (36\%) not executable. These results reveal that while artifact sharing has increased, a significant fraction still requires substantial improvement to meet execution standards.
The data suggests that growth in artifact sharing does not necessarily translate to improved executability. Without targeted community efforts to specify the environment, document, and establish packaging standards, the executability goals of open science will remain elusive.

\noindent\textbf{Level of Effort Required to Execute Replication Packages.}
Figure~\ref{fig:RQ1_effort_mod}(a) presents the relationship between executability status and the level of effort required to execute the artifacts. The results show a stark contrast in effort distribution across executable, partially executable, and non-executable artifacts.
Among the 40 executable artifacts, the majority (72.5\%) required either low (11) or moderate (18) effort. However, 11 executable artifacts (27.5\%) still demanded high effort, indicating that even runnable artifacts often lack sufficient documentation, packaging, or automation support.
% quality.
The situation is more difficult for partially executable artifacts. Among the 24 partially executable artifacts, 13 required moderate, and 11 required high effort, yet none could be executed successfully. 
Even for the non-executable artifacts (36), we invested non-trivial time: 52.78\% (19/36) required moderate-to-high effort before reaching a dead end. The remaining 47.22\% (17/36) failed early due to severe deficiencies or incompleteness and were classified under low effort, not because they were simple, but because they were unworkable from the start. These results emphasize the urgent need for community-wide practices that minimize unnecessary overhead by promoting thorough environment specification, proper automation, and quality assurance of replication packages.

\begin{figure}[htb]
\centering
\includegraphics[width=0.90\columnwidth]{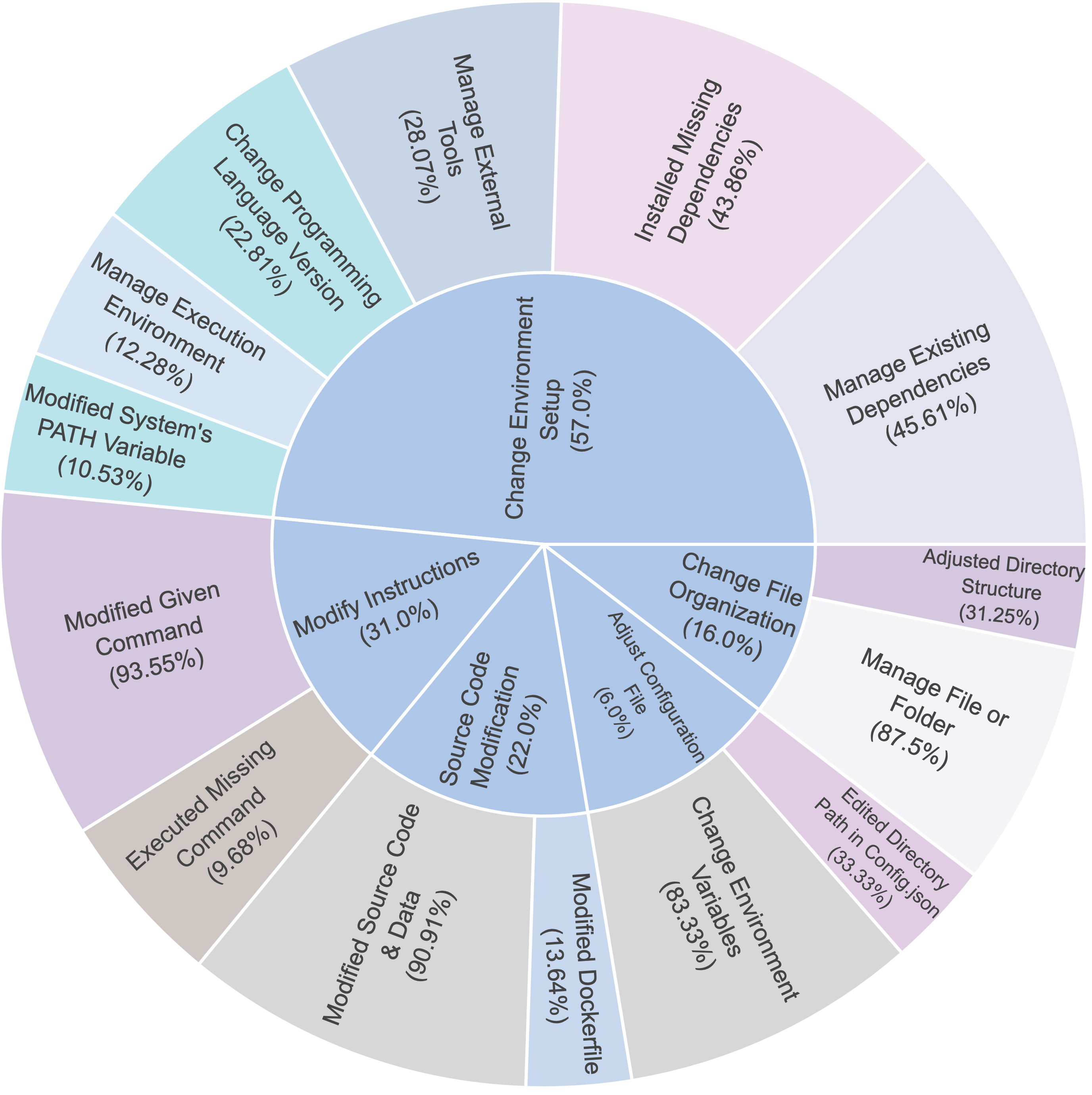}
\caption{Types of Modifications Required for Execution}
\Description{A hierarchical taxonomy chart showing types of modifications required for artifact execution. The chart has 5 main categories at the top level: Configuration, Dependency, Code, Environment, and Documentation.}
\label{fig:RQ2_modifications}
\vspace{-3mm}
\end{figure}

\noindent\textbf{Level of Modification Required to Execute Replication Packages.}
Figure~\ref{fig:RQ1_effort_mod}(b) visualizes the extent of modifications needed across artifacts with varying executability statuses. 
Among the 40 executable artifacts, only 13 (32.5\%) worked without any modification. An additional 6 artifacts (15\%) required only minor changes, such as adjusting directory paths or fixing broken shell commands. However, a significant portion of the executable artifacts (37.5\%) required major-to-critical modifications, including editing source code or resolving complex environment conflicts, to achieve successful execution. Such findings challenge the assumption that ``executable'' implies low effort.

The situation deteriorates for partially executable artifacts. Of the 24 analyzed, 79.17\% (19/24) required major or critical modifications. These typically involved deep architectural issues, undocumented assumptions, or persistent runtime errors, making full reproduction infeasible despite extensive manual intervention.
Among 36 non-executable artifacts, 20 (55.56\%) underwent major or critical modifications before ultimately failing. This indicates that significant effort was still invested in troubleshooting, debugging, and altering code, but the artifacts could not be salvaged. The remaining 16 artifacts (44.44\%) required no or minor modification, not because they were simple, but because execution attempts were abandoned early due to insurmountable issues such as missing datasets or unusable documentation.

Overall, these results suggest that artifact sharing alone is insufficient. Without proper environmental details, dependency control, and clear code, artifacts often remain unusable. The high modification burden highlights a pressing need for stronger tooling, standards, and author-side validation.

\subsection{Answering RQ2: Types of Modifications Needed to Execute Replication Packages}
\label{subsec:findings-rq2}

We conducted a qualitative analysis (as discussed in Section \ref{subsub:modification-types}) and labeled the observed modifications using a hierarchical classification spanning five broad categories and 14 specific types.

Figure~\ref{fig:RQ2_modifications} shows that environment-related issues were most prevalent, affecting 57\% of artifacts. Common problems included installing missing dependencies (43.86\%) and resolving version conflicts (45.61\%), often due to unpinned dependencies or outdated instructions. In one case, resolving a circular Python dependency took nearly 10 hours, which could be avoided with proper versioning.
Instruction-related changes occurred in 31 artifacts, with 92.86\% of execution commands requiring adjustments, such as fixing syntax, flags, or paths. A notable case involved a five-hour debugging effort due to a missing \textit{--gpus} flag in a Docker command, resulting in a misleading “division by zero” error.

Code-level modifications were required in 22\% of the artifacts, primarily to address runtime errors (90.91\%). These were particularly challenging due to poor code documentation, as only 26.67\% (4/15) of such artifacts were ultimately executed, often requiring substantial effort. One required modifying the \texttt{tensor2tensor} library logic twice, highlighting a broader lack of maintainability.
The remaining 22\% of artifacts required surface-level changes. Specifically, 16\% needed folder restructuring to fix broken paths, and 6\% required environment variable or path adjustments in config files. Although simple, such issues were common and indicated inadequate quality control in artifact preparation.

\begin{figure}[htb]
\centering
\includegraphics[width=0.90\columnwidth]{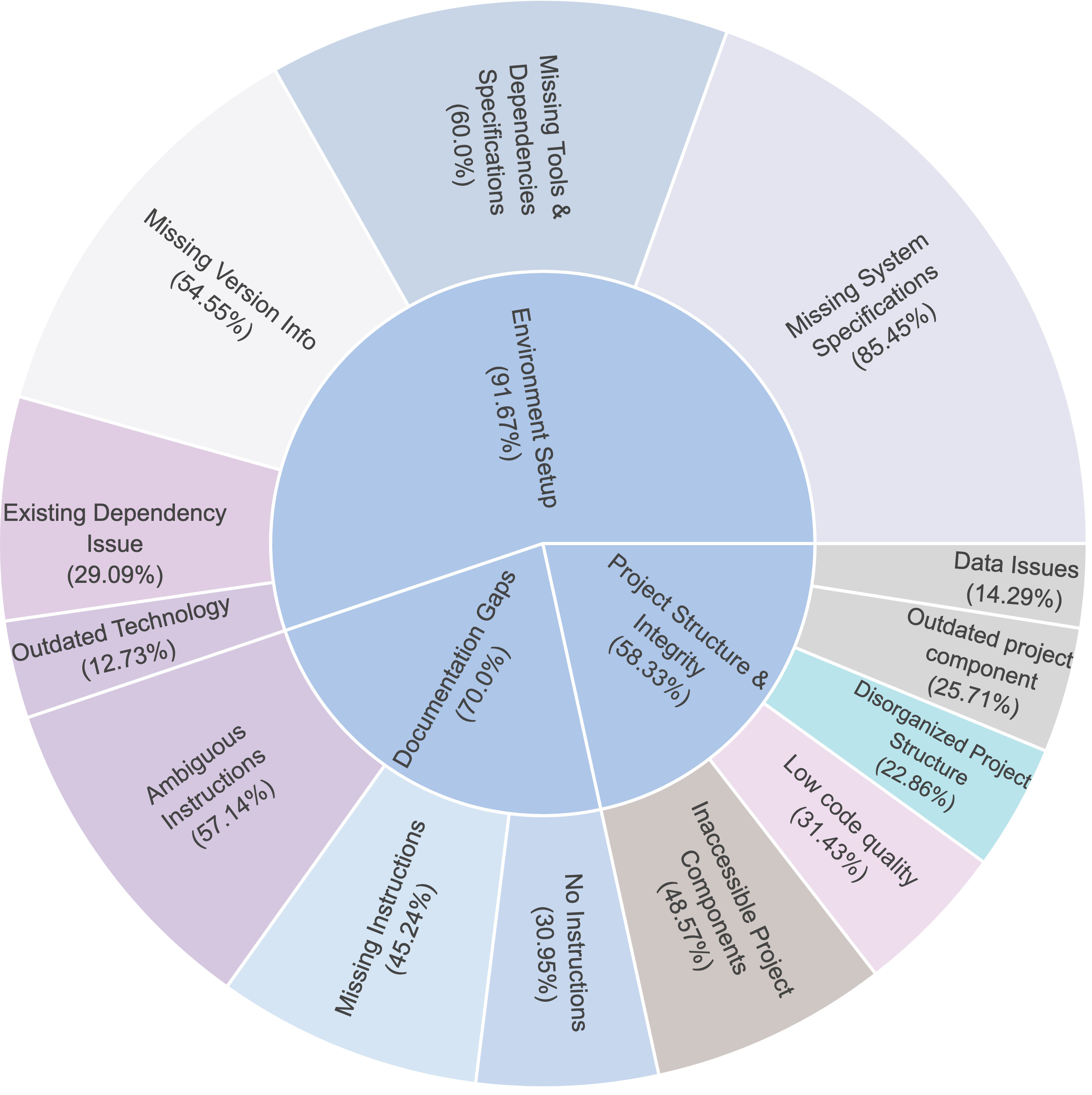}
\caption{List of Challenges Preventing Execution}
\Description{A hierarchical chart displaying execution barriers organized into 3 main themes: Documentation Issues, Environment Setup Challenges, and Technical Issues. Each theme branches into specific challenge types, with frequency counts provided.}
\label{fig:RQ3_challenges}
\end{figure}

\subsection{Answering RQ3: Challenges that Prevent Executability of Replication Packages}
\label{subsec:findings-rq3}

Figure~\ref{fig:RQ3_challenges} presents the major barriers to execution, along with their frequency and percentage among the artifacts marked as ``Not Executable'' or ``Partially Executable.''

Similar to the modification patterns, environment setup challenges emerged as the most frequent issue, affecting 91.67\% of the artifacts with execution failures. Within this category, 85.45\% of the artifacts lacked essential system specifications. These omissions included missing operating system details, platform-specific configurations, or hardware-related requirements such as GPU support, RAM, or disk space. In 60\% of the artifacts, dependency information, such as required libraries or external tools, was not documented. Furthermore, 54.55\% failed to mention the versions of the programming languages or dependencies used in the project (e.g., Python, Java), making it difficult to align our environment with that of the original experiment. While any one of these issues alone might not entirely prevent execution, their combination created significant barriers. For instance, 35\% of the artifacts were still executable when only one of these issues was present. However, for artifacts that faced all three challenges simultaneously, the executability rate dropped sharply to 8\%, with more than half (52\%) failing to run. In contrast, when none of these three environment-related issues were present, 78.8\% of the artifacts executed successfully, underscoring the importance of providing clear and complete environment setup instructions in replication packages.

The second major challenge category was inadequate documentation, affecting 70\% of non-executable or partially executable artifacts. Within this group, we observed three recurring problems. First, in 57.1\% of the artifacts, the documentation was ambiguous, where steps were either unclear or not logically ordered, making it difficult to follow. Second, 45.2\% of the artifacts had missing instructions, with key steps in the execution process entirely omitted. While we were sometimes able to overcome these issues using our technical expertise and by inferring missing details from the provided materials, success was inconsistent. When only one of these documentation issues was present, we could still execute 42.4\% of the artifacts. However, when both ambiguous and missing documentation occurred together, the executability rate dropped to 7.7\%, with 61.5\% of those cases ending in partial execution. The situation was especially difficult when artifacts lacked any usable documentation. In this subset categorized as "No Documentation", we were able to execute only 7.1\% of the artifacts, whereas the remaining 92.9\% could not be executed.

The final set of challenges related to inaccessible or outdated components in the replication packages. For artifacts with missing project elements such as datasets, scripts, or configuration files, only 15\% could be executed. Success in these few cases was achieved either because the authors provided a Docker image alongside the codebase or because we reconstructed missing files using clues from other resources. Outdated components were another common obstacle. In these cases, changes were made to parts of the repository while other dependent components were left untouched, leading to inconsistencies and execution errors. Only 18.2\% of these artifacts were executable, while the majority (63.6\%) could only be partially executed due to unresolved conflicts or deprecated configurations.

These findings indicate that while many artifacts are shared with the intention of supporting reuse, a large proportion fall short due to avoidable issues such as incomplete documentation, missing configuration details, or outdated codebases. Addressing these challenges through clearer documentation, robust environment specifications, and well-maintained artifact repositories would significantly improve the executability of shared artifacts and move the SE community closer to the goals of open science.

\subsection{Answering RQ4: Reproducibility Outcomes and Contributing Factors}
\label{subsec:findings-rq4}

Among the 100 replication packages sampled in our study, 64 were either fully or partially executable and thus qualified for further evaluation of result reproducibility. 

\begin{figure}[htb]
\centering
\includegraphics[width=0.95\columnwidth]{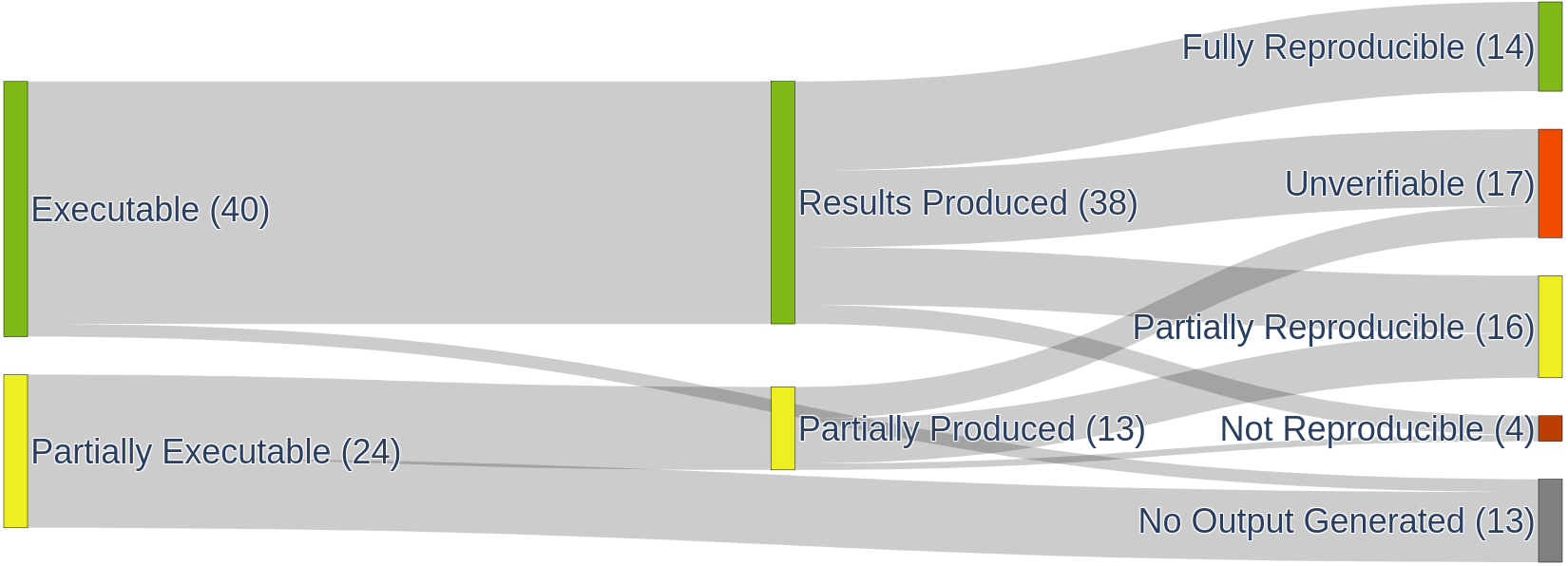}
\caption{Reproducibility Status of Executable Artifacts}
\Description{A stacked bar chart showing reproducibility results for 64 executable artifacts. The chart is divided into three colored segments: fully reproducible (14 artifacts, 14\%), partially reproducible (16 artifacts, 16\%), and not reproducible (34 artifacts, 70\%).}
\label{fig:RQ4_reproducibility}
\vspace{-3mm}
\end{figure}

Figure~\ref{fig:RQ4_reproducibility} shows the distribution of reproducibility outcomes. While RQ1 revealed challenges in executability, reproducibility results are even more concerning. Only 8\% of artifacts were fully reproducible without modification, 5\% with minor adjustments, and 1\% with major effort, yielding a total reproducibility rate of just 14\%. Another 16\% were partially reproducible, often requiring major or critical modifications (62.5\%). These results emphasize that successful execution does not ensure reproducibility. Fully reproducible artifacts were better documented and required less effort (85.71\% low to moderate), whereas 50\% of partially reproducible artifacts required high effort.

Among the 40 fully executable artifacts, only 35\% reproduced results matching the original study.
Another 30\% produced divergent outputs, and 5\% produced no results. The situation was worse for partially executable artifacts, where 45.83\% failed to generate any usable results due to incomplete pipelines, missing components, runtime crashes, or logical errors.
A major barrier was the absence of validation guidance. For 30\% of executable artifacts and 20.8\% of partially executable artifacts, we could not verify the results despite execution due to missing instructions or unclear output structures. Only 6.3\% provided sufficient validation steps for comparison.

Overall, reproducibility remains a major gap. Running an artifact does not guarantee credible or interpretable outcomes. Researchers must go beyond code and data sharing to ensure proper documentation, output clarity, and result validation to enable reproducible SE research.

\section{Discussion} 
In this section, we present actionable guidelines to improve the executability and reproducibility of replication packages, evaluate our own package against these guidelines, and discuss the broader implications of our study.

\subsection{Actionable Guidelines for Improving Executability}
\label{subsec:guidelines}

\noindent \textbf{G1: Provide Comprehensive and Structured Documentation.}
Findings from RQ2 and RQ3 reveal that well-structured documentation is one of the key determinants of artifact executability and reproducibility. Our analysis identified five essential documentation components:

\textbf{(C1) Project Metadata} -- concise overview of the project purpose and materials (e.g., scripts, datasets, models).

\textbf{(C2) System Specifications} -- details of OS, hardware (e.g., GPU) requirements, and external tools, each with version information.

\textbf{(C3) Installation \& Environment Setup} -- programming language, dependencies, and essential package versions. If any packages are used from a local registry or built from source, include clear instructions for setting them up.

\textbf{(C4) Execution Steps} -- entry points, commands, parameters, and expected outputs.

\textbf{(C5) Validation Steps} -- clear instructions for verifying results, including scripts, outputs, and links to paper figures or tables.

\noindent According to our analysis, executability and reproducibility increased sharply with the inclusion of these components. Artifacts lacking all five components failed in all cases to execute or reproduce, while those with two components achieved 37.5\% executability and 55.6\% reproducibility. Artifacts containing all five reached 85.7\% executability and 83.3\% reproducibility, highlighting the decisive impact of comprehensive documentation.

\begin{snugshade}
\noindent\textbf{\textit{Summary of G1:}} Comprehensive, structured documentation is essential for making artifacts executable and reproducible. Including all five components turns artifacts into transparent, verifiable, and reusable research assets that advance open science in SE.
\end{snugshade}

\noindent \textbf{G2: Avoid Configuration Leakage.} During our execution, several artifacts failed even when we carefully followed the provided documentation. In one case, a \texttt{requirements.txt} file listed more than 100 Python packages, including local path references and packages built from source. We removed one unnecessary package that caused errors, and after resolving the dependency conflicts, it was not reinstalled even as a transitive dependency.
This file appeared to capture the author’s entire local environment rather than the specific dependencies required to run the project. After replacing local and source builds with their official releases and resolving version conflicts, the environment was successfully resolved.
This confirmed that the original failure stemmed from implicit dependency bindings rather than faulty implementation.

Such issues reflect configuration leakage from the development environment, where artifacts depend on implicit settings, cached tools, or residual files from the author's machine that are not explicitly documented. To avoid this problem, authors should test their replication packages in a clean, isolated environment before submission to identify hidden dependencies, missing paths, and unintended configuration carryovers.

\begin{snugshade}
\noindent\textbf{\textit{Summary of G2:}} Testing artifacts in a clean environment helps detect configuration leakage early and ensures that they remain portable, executable, and independent of the author's local setup.
\end{snugshade}

\noindent \textbf{G3: Combine Containerized Environments with Accessible Source Code.}
Containerization encapsulates the development environment and dependencies, improving portability and reducing setup complexity, while source code ensures transparency, flexibility, and long-term adaptability.
In our dataset, 34 artifacts included both containerized environments (e.g., Docker, VM) and raw source code.
Artifacts from containerized setups achieved higher executability (67.47\%) and reproducibility (52.17\%) than source-only artifacts (31\% and 33.25\%, respectively). Most containerized artifacts (83.33\%) ran successfully with low to moderate effort, and 75\% required no modification, indicating that encapsulated environments substantially simplify execution. This becomes particularly critical when executing a study that requires complex setup procedures and is highly sensitive to environmental variations.
However, containerization introduces trade-offs. When execution failed, 57.14\% of containerized artifacts were unrecoverable due to restricted access or opaque configurations. In contrast, raw source code, though more demanding to configure, offers greater transparency for debugging, reuse, and extension.

\begin{snugshade}
\noindent\textbf{\textit{Summary of G3:}} Providing both containerized environments and source code offers a sustainable approach to artifact sharing, as containers lower the barrier to execution while source code preserves adaptability, reusability, and transparency.
\end{snugshade}

\subsection{Preparation and Evaluation of Our Replication Package}
\label{subsec:assessment-of-replication-package}

\noindent \textbf{Preparation of our Replication Package.}
Our replication package was constructed following the actionable guidelines derived from executing and reproducing 100 ICSE replication packages in this study. It is organized into two complementary parts. 
\underline{Part 1} contains all analysis scripts and datasets required to regenerate the results reported in this paper. It was developed in full compliance with the documentation and packaging guidelines introduced in Section \ref{subsec:guidelines}. Along with the source code, we also provide a \texttt{Dockerfile} to run our experiments in an isolated environment.
\underline{Part 2} provides a curated, per-artifact record of our execution and reproduction attempts for 100 ICSE replication packages. It does not modify the original artifacts but documents our interactions with them to ensure full traceability and transparency. Each artifact record includes four files (\texttt{README.md}, \texttt{execution-log.md}, \texttt{patch.md}, and \texttt{updated-instructions.md}) and one folder (\texttt{/errors}) when applicable. The detailed purpose of these files can be found in the \texttt{README.md} file of our replication package \cite{ReplicationPackage}.

\noindent \textbf{Evaluation of our Replication Package.}
To assess the credibility and usability of our replication package, we recruited \textit{four} independent evaluators. Two of them have over eight years of professional software development experience, and two are software engineering researchers with active publication records in empirical and reproducibility studies. This combination ensured both practical and scientific perspectives on the evaluation.

$\bullet$ \textbf{\textit{Assessment Summary (Part One)}.} All four evaluators executed part one in full using both the Jupyter notebooks and the Docker setup, following the provided instructions. 
They evaluated documentation completeness, executability, reproducibility, and overall usability in accordance with our proposed guidelines.
The evaluators unanimously confirmed that the package contained all five essential documentation components (\emph{Project Metadata, System Specification, Installation \& Environment Setup, Execution Steps,} and \emph{Validation Steps}), thereby fully satisfying G1 (comprehensive documentation). 
For G2 (avoiding configuration leakage), all evaluators successfully reproduced the results without any modifications or environment-related issues. To further strengthen this assessment, the first author independently validated Part One in clean, isolated environments across multiple operating systems, confirming the absence of hidden dependencies or environment leakage.

Finally, in accordance with G3 (combining Docker with accessible source code), both the source code and Docker setup were provided with separate instructions, and all evaluators successfully reproduced the results using both approaches, demonstrating that the setup is robust, portable, and accessible to users with different preferences and technical backgrounds.

$\bullet$ \textbf{\textit{Assessment Summary (Part Two)}.} Table \ref{tab:part2-eval} summarizes the assessment of Part II by the four evaluators (individual responses are available in our replication package \cite{ReplicationPackage}). We evaluated this part using a stratified random sample of 20 artifacts drawn proportionally from the three executability categories (8~Executable, 5~Partially~Executable, and 7~Not~Executable) and distributed evenly among four evaluators. 
Each evaluator first reviewed the \texttt{README.md} file to verify the factual accuracy of our documentation. They then executed each artifact twice: once strictly following the original instructions and once using our meta-documentation (\texttt{execution-log.md}, \texttt{patch.md}, and \texttt{updated-instructions.md}). This two-pass execution allowed evaluators to assess both the baseline usability of the artifacts and the effectiveness of our structured documentation in improving comprehensibility and execution success.

\begin{table}[t]
\centering
\caption{External Evaluation Results for Part Two (E = Executable, count = 8; PE = Partially Executable, count = 5; NE = Not Executable, count = 7; T = Total, count = 20)}
\label{tab:part2-eval}
\resizebox{3in}{!}{
\begin{tabular}{l l c c c c}
\hline
\textbf{Criterion} & \textbf{Judgment} 
& \textbf{E} 
& \textbf{PE} 
& \textbf{NE} 
& \textbf{T} \\
\hline
\hline
\multirow{2}{*}{Factual Accuracy} 
  & Accurate (As is)               & 7 & 4 & 7 & 18 \\ %\crule(lr){2-6}
  & Minor modifications      & 1 & 1 & 0 & 2 \\
\hline
\multirow{2}{*}{\makecell[l]{Execution Steps\\(Ours vs. Original)}}
  & Ours Better             & 3 & 5 & 4 & 12 \\ %\crule(lr){2-6}
  & Both Same               & 5 & 0 & 3 & 8 \\
\hline
\multirow{3}{*}{Patch Correctness} 
  & Worked        & 3 & 5 & 3 & 11 \\ %\crule(lr){2-6}
  & Did not Work        & 1 & 0 & 0 & 1 \\ %\crule(lr){2-6}
  & No Patch Provided    & 4 & 0 & 4 & 8 \\
\hline
\multirow{2}{*}{Usability} 
  & Improved                & 3 & 5 & 4 & 12 \\ %\crule(lr){2-6}
  & Unchanged               & 5 & 0 & 3 & 8 \\
\hline
\multirow{2}{*}{Agreement Level} 
  & Agreed                  & 7 & 5 & 7 & 19 \\ %\crule(lr){2-6}
  & Disagreed               & 1 & 0 & 0 & 1 \\
\hline
Final Status 
  & Unchanged               & 8 & 5 & 7 & 20 \\
\hline
\end{tabular}}
% \vspace{-2mm}
\end{table}

All evaluators reproduced our reported results, and the final status of executability and reproducibility remained unchanged across all 20 artifacts. Nineteen artifacts were fully consistent with our recorded classifications. 
The single disagreement occurred because one evaluator's local machine already contained a dependency (\texttt{build-essential}) that our clean Docker setup required to be installed explicitly.
Evaluators rated 18 artifacts as factually accurate and 2 as partially accurate, noting only minor corrections that were later re-checked by the authors. In 12 cases, our meta-documentation was judged to be clearer and more usable than the original instructions. Usability remained unchanged for 5 artifacts whose original documentation was already sufficient. The remaining 3 artifacts were not executable, and their replication packages contained no execution instructions, so we could not further improve their execution processes.

Although we could not directly apply our proposed guidelines (Section~\ref{subsec:guidelines}) to third-party artifacts, we compensated by providing structured meta-documentation, which improved traceability and reproducibility. In line with \textbf{G1}, we evaluated each artifact against the five documentation components and filled missing details in our meta-files, which evaluators later confirmed as factually accurate. For \textbf{G2}, since we could not control environment leakage in external packages, we mitigated it by recording complete execution logs, patches, and errors and providing them in our replication package in a structured, hyperlinked format. Regarding \textbf{G3}, re-containerizing all 100 heterogeneous artifacts was infeasible. Instead, we documented key environmental and configuration details that evaluators reported as helpful for improving clarity and reusability, except in cases where the original packages lacked sufficient information for full replication.

Given that our replication package consists of two interrelated parts, evaluating the main \texttt{README.md} file was essential to ensure that users could understand the study design, the roles of each part, and the steps for reusing our artifact. Evaluators assessed it for documentation quality, inter-component linkage, and ease of navigation, awarding a perfect average score of 5/5 for clarity and completeness. Minor suggestions were carefully incorporated into the final version of our replication package. Overall, the combination of industry and academic evaluators provided balanced evidence that our replication package is usable in practice, transparent in structure, and methodologically sound for research reuse.

\subsection{Implications}
\label{subsec:implications}

Our study provides actionable insights for multiple stakeholders in the SE community, including researchers, conference organizers, and industry practitioners.
For \textbf{Researchers}, the findings of this study can help develop more usable and reproducible replication packages. Researchers can follow our proposed documentation structure to ensure clear execution and validation paths, and they can provide both source code and containerized environments to balance usability and extensibility. They can also apply practical strategies derived from our experience, such as using version-buffer heuristics when version details are missing. For \textbf{Conference Organizers}, our results can inform the development of improved artifact submission and evaluation policies. For example, asking reviewers to validate packages in isolated environments (e.g., fresh Docker containers) can reveal hidden configuration issues and more accurately reflect real-world usability. For the \textbf{SE Community}, the shortcomings we identified underscore the need for dedicated platforms to host and maintain research artifacts, ideally with automated quality checks that detect common barriers to executability and reproducibility and guide authors in following best practices to advance open science in SE research.

\section{Threats to Validity}

Threats to \emph{external validity} refer to the extent to which our findings generalize beyond the studied sample. Our dataset comprises replication packages from ICSE research-track papers, which may limit its generalizability to other SE venues or domains.
However, ICSE is one of the most prestigious SE conferences, with rigorous artifact evaluation processes, a diverse range of SE topics, and consistently high-quality publications.
As a result, our findings likely represent a best-case scenario for open science practices in the field. This study can be extended to other major venues, such as FSE and ASE, to examine whether similar patterns appear across the broader SE research community.
A second external validity threat concerns temporal generalizability, as we examine a 10-year window (2015-2024). This period is broad enough to capture evolving artifact practices but may not reflect earlier trends. Expanding the window further would dilute yearly samples and reduce representativeness. To mitigate this, we applied stratified random sampling across years to ensure balanced coverage.

Threats to \textit{internal validity} relate to potential errors in data collection and analysis. Our evaluation involved several manual steps, including artifact identification, executability assessment, effort classification, and reproducibility evaluation. 
To reduce human error, we employed a multi-pass review process, and ambiguous cases were independently examined by multiple authors to reach consensus.
Evaluator expertise may influence executability outcomes. Although two experienced authors independently attempted execution while strictly adhering to the original documentation, differences in technical background, tool familiarity, or system configurations may lead other practitioners to obtain different results. Thus, artifacts classified as non-executable should be interpreted as unsuccessful under controlled and informed conditions rather than as evidence of absolute infeasibility.

Threats to \textit{construct validity} concern the correctness and consistency of how we measured executability, effort, and reproducibility. While these constructs are commonly used in empirical studies, some steps, such as identifying modification types or diagnosing failure causes, require judgment. We mitigated subjectivity bias through structured coding schemes, predefined criteria, and agreement across evaluators. When one evaluator could not execute an artifact, a second evaluator repeated the process independently. If both encountered the same issue, we attributed the failure to the artifact. Our effort estimates, based on agile-style bucketed time ranges, may be sensitive to changes in the boundaries. A sensitivity analysis showed that increasing the ranges by 5-10 percent produced no changes, while decreasing them caused only minor shifts, such as a few low-effort cases moving to moderate, indicating a negligible impact on overall findings. Hardware constraints present another construct validity threat. Some artifacts required extremely high-end setups, such as machines with 250 GB of RAM or multiple NVIDIA A100 GPUs, which we could not precisely replicate. We mitigated this limitation by using the closest available hardware (Section~\ref{sec:experimental-setup}) and adjusting permitted configurations, such as batch size and dependency versions, accordingly.

\section{Related Work}

Open science has gained increasing traction in SE, with several studies promoting it as a foundation for rigorous research, empirical investigations, and technical and methodological solutions to improve research transparency and reproducibility. \citet{OliveiraJr2024} proposed a conceptual framework that treats open science as a requirement for promoting a cultural shift within the SE community. \citet{Cordeiro2025b} examined how reproducibility is defined, evaluated, and supported, while others highlight reproducible research as a scientific standard for addressing validity concerns in SE \cite{Madeyski2017, Tkachenko2025}.

Prior studies have also assessed the reproducibility of SE research to identify existing practices, challenges, and gaps. \citet{Liu2024} report an upward trend in artifact sharing but caution that availability alone does not guarantee usability. Similarly, \citet{Cordeiro2025a}, based on a survey of SE researchers, found that limited data sharing and incomplete experimental details remain key barriers to reproducibility. Beyond these issues, several technical factors may further hinder reproducibility, including dependency problems, incomplete documentation, and insufficient information to reproduce results \cite{Ajayi2023, Bonneel2020, RodriguezPerez2018}. In deep learning-based SE studies, additional challenges arise from inconsistent evaluation practices, unstable experimental setups, and missing details about baselines, preprocessing steps, and hyperparameters \cite{Nong2023, Liu2022, Mukhtar2024}.

Researchers have also explored technical and methodological solutions to improve the availability and reproducibility of SE research. \citet{Mendez2019} introduced an open science initiative for the Empirical Software Engineering journal, while \citet{GonzalezBarahona2023} proposed structured assessment protocols for Mining Software Repository studies. \citet{Mahmood2018} further recommended improved reporting guidelines and incentive mechanisms to strengthen replication efforts. Additional technical solutions include containerization to lock execution environments \cite{Costa2024} and modeling environment variability to identify potential sources of irreproducibility \cite{Acher2024}.

% Research Gap.
While prior work has assessed artifact availability and advocated better practices, a comprehensive evaluation of executability, required effort, modification types, and actual reproducibility remains an unmet attempt in SE research. Our study fills this critical gap by empirically analyzing 100 ICSE artifacts, quantifying usability barriers, and providing actionable recommendations. 
% This section synthesizes the most relevant work aligned with our investigation.

\section{Conclusion}

Despite the growing emphasis on open science in SE research, the practical usability of shared artifacts has received insufficient attention. 
In this study, we provide a comprehensive empirical assessment of executability and reproducibility across 100 ICSE replication packages. 
By investing roughly 650 person-hours of analysis, we uncovered substantial gaps. Only 40\% of artifacts were executable, about one-third of these ran without modification, and reproducibility was achieved in only 35\% of the executable artifacts.
To help close this gap, we proposed three actionable guidelines: adopt comprehensive documentation, validate artifacts in clean environments to avoid configuration leakage, and supply both containerized setups and accessible source code (where applicable). Our findings show that artifacts that include these components achieve substantially higher levels of executability and reproducibility.
Strengthening open science in SE requires not only sharing artifacts but ensuring they are runnable, verifiable, and usable by others. Following the guidelines introduced in this study can help researchers and conference organizers advance reproducible and trustworthy SE research.

\begin{acks}
This research is supported in part by the Natural Sciences and Engineering Research Council of Canada (NSERC) Discovery Grants program, the Canada Foundation for Innovation's John R. Evans Leaders Fund (CFI-JELF), and by the industry-stream NSERC CREATE in Software Analytics Research (SOAR).
\end{acks}

\balance
\bibliographystyle{ACM-Reference-Format}
\bibliography{references}

\end{document}